\documentclass{IEEEtran}

\usepackage{amssymb}
\usepackage{makeidx}         
\usepackage{amsmath}
\usepackage{multicol}        
\usepackage[bottom]{footmisc}
\usepackage{graphicx}
\usepackage{color}
\usepackage{amsfonts}
\usepackage[latin1]{inputenc}
\usepackage{epstopdf}

\newcommand{\eq}{\begin{equation}}
\newcommand{\eeq}{\end{equation}}
\newcommand{\eqn}{\begin{eqnarray}}
\newcommand{\eeqn}{\end{eqnarray}}

\newcommand{\bsea}{\begin{subeqnarray}}
\newcommand{\esea}{\end{subeqnarray}}
\newcommand{\nn}{\nonumber}

\newcommand{\Sp}[2]{\left< #1,#2 \right> }

\newcommand{\Min}[1]{\,\underset{#1}{\mathrm{min}}\,}

\newcommand{\Inf}[1]{\,\underset{#1}{\mathrm{inf}}\,}

\newcommand{\Span}[1]{\mathrm{span}\left\{ #1\right\}}

\newcommand{\Range}{\mathop{\rm Range}}
\newcommand{\de}{\mathrm{d}}
\newcommand{\tr}{\mathop{\rm tr}}  
\newcommand{\e}[1]{\mathrm{e}^{#1}}
\newcommand{\Set}[1]{\left\{ #1\right\}}

\newcommand{\Dc}{ \mathcal{D}}

\newcommand{\Hc}{ \mathcal{H}}
\newcommand{\Ic}{ \mathcal{I}}

\newcommand{\Kc}{ \mathcal{K}}
\newcommand{\Lc}{ \mathcal{L}}
\newcommand{\Mc}{ \mathcal{M}}
\newcommand{\Nc}{ \mathcal{N}}

\newcommand{\Sc}{ \mathcal{S}}

\newcommand{\Uc}{ \mathcal{U}}

\newcommand{\Cs}{ \mathbb{C}}

\newcommand{\Ns}{ \mathbb{N}}

\newcommand{\Rs}{ \mathbb{R}}
\newcommand{\Ss}{ \mathbb{S}}

\newcommand{\mz}{\color{black}}

\def\qed{\hfill \vrule height 7pt width 7pt depth 0pt \smallskip}

\newcounter{pippo}

\newtheorem{remark}{Remark}[section]
\newtheorem{teor}{Theorem}[section]
\newtheorem{corr}{Corollary}[section]
\newtheorem{propo}{Proposition}[section]
\newtheorem{lemm}{Lemma}[section]
\newtheorem{exam}{Example}
\newtheorem{probl}[pippo]{Problem}
\newtheorem{defn}{Definition}[section]
\newcommand{\teo}{\begin{teor}}
\newcommand{\eteo}{\end{teor}}
\newcommand{\cor}{\begin{corr}}
\newcommand{\ecor}{\end{corr}}
\newcommand{\prop}{\begin{propo}}
\newcommand{\eprop}{\end{propo}}
\newcommand{\lem}{\begin{lemm}}
\newcommand{\elem}{\end{lemm}}
\newcommand{\ex}{\begin{exam}}
\newcommand{\eex}{\end{exam}}
\newcommand{\pb}{\begin{probl}}
\newcommand{\epb}{\end{probl}}
\newcommand{\df}{\begin{defn}}
\newcommand{\edf}{\end{defn}}
\newcommand{\aprop}{\begin{apropo}}
\newcommand{\eaprop}{\end{apropo}}
\newcommand{\alem}{\begin{alemm}}
\newcommand{\ealem}{\end{alemm}}
\newcommand{\rem}{\begin{remark}}
\newcommand{\erem}{\end{remark}}

\begin{document}
\title{Minimum Relative Entropy for Quantum Estimation: Feasibility and General Solution}


\author{Mattia~Zorzi, Francesco~Ticozzi and Augusto~Ferrante\thanks{Work partially supported by the Italian Ministry for Education and Research (MIUR) under PRIN grant n. 20085FFJ2Z
``New Algorithms and Applications of System Identification and Adaptive Control", and by the University of Padua under the QUINTET project of the Dept. of Information Engineering and  the QFUTURE strategic Project.}
\thanks{M. Zorzi and A. Ferrante are with the
Dipartimento di Ingegneria dell'Informazione, Universit\`a di
Padova, via Gradenigo 6/B, 35131 Padova, Italy ({\tt\small
zorzimat@dei.unipd.it},{\tt\small augusto@dei.unipd.it}), and F. Ticozzi is with with the
Dipartimento di Ingegneria dell'Informazione, Universit\`a di
Padova, via Gradenigo 6/B, 35131 Padova, Italy ({\tt\small ticozzi@dei.unipd.it}),
 and the Deptartment of Physics and Astronomy, Dartmouth College, 3127 Wilder, Hanover, NH (USA).}}

\markboth{DRAFT}{Shell \MakeLowercase{\textit{et al.}}: Bare Demo of IEEEtran.cls for Journals}

\maketitle

\begin{abstract}
We propose a general framework for solving quantum state estimation problems using the minimum relative entropy criterion. A convex optimization approach allows us to decide the feasibility of the problem given the data and, whenever necessary, to relax the constraints in order to allow for a physically admissible solution. Building on these results, the variational analysis can be completed ensuring existence and uniqueness of the optimum. The latter can then be computed by standard, efficient standard algorithms for convex optimization, without resorting to approximate methods or restrictive assumptions on its rank.
\end{abstract}


\section{Introduction}
Quantum devices implementing information processing tasks promise
potential advantages with respect to their classical counterparts
in a remarkably wide spectrum of applications, ranging from secure
communications to simulators of large scale physical systems
\cite{feynman-QC,zeilinger,nielsen-chuang}.

In order to exploit quantum features to the advantage of a desired task, tremendous challenges are posed to
experimentalists and engineers, and many of these have
stimulated substantial theoretically-oriented research. Which
particular problem is critical depends on
the physical system under consideration: from optical integrated
circuits to solid-state devices, the tasks in the device
engineering, protection from noise and control are manifold
\cite{nielsen-chuang, zeilinger, dalessandro-book,
wiseman-book,altafini-tutorial}.
However, {\em quantum estimation}
\cite{PARIS_QSTATE_ESTIMATION_2004} problems are ubiquitous in
applications, be it in testing the output of a quantum algorithm,
in reconstructing the behavior of a quantum channel or in
retrieving information at the receiver of a communication system
\cite{nielsen-chuang,petz-book,ZORZI_MINIMAL_RESOURCES_2011,lidar-estimation,pino-tomography,villoresi}.
In this paper we focus on state estimation problem for
finite-dimensional quantum system, namely the reconstruction of a
trace-one, positive {\mz semidefinite} matrix given from data, and
in particular on an estimation method that addresses two critical
problems for most real-world situations. The first arises when
only a small set of potentially noisy data is available, yielding
no physically-acceptable solution; the second regards situation in
which the system dimension is large and the available measurement
are insufficient to completely determine the state.

To address this second issue, a typical approach in both the
classical  and the quantum world is to resort to a MAXENT principle
\cite{JAYNES106,JAYNES108,burg1975maximum,petz-entropy,PARIS_QSTATE_ESTIMATION_2004,ziman-maxent},
where one opts for a ``maximum ignorance'' criterion on the choice
of parameters that are not uniquely determined by data. The MAXENT
estimation can indeed be seen as a particular case of {\em minimum
relative entropy} estimation, \cite{Shore}, where the
information-theoretic pseudo-distance of the estimated state with
respect to some {\em a priori} state is minimized subjected to a set
of constraints representing the available data
\cite{RELATIVE_ENTROPY_GEORGIOU_2006,vedral-relativeentropy,PARIS_MIN_KL_2007,pavon-discretebridges,ticozzi-timereversal}.
This {\em a priori} information introduces a new ingredient with
respect to typical maximum-likelihood methods for quantum estimation
\cite{PARIS_QSTATE_ESTIMATION_2004,ZORZI_MINIMAL_RESOURCES_2011}.

A quantum minimum relative entropy method for state estimation has been discussed in \cite{PARIS_MIN_KL_2007,BRAUNSTEIN_GEOMETRY_INFERENCE_1996}, where {\em approximate solutions} to minimum relative entropy problems are provided: the estimates are shown to be good approximation of the optimal solution when this is close enough to the prior. On the other hand, a way towards the computation of the exact {\em optimal solution} is indicated in \cite{RELATIVE_ENTROPY_GEORGIOU_2006}: Georgiou has analyzed
the MAXENT problem for estimating positive definite matrices, providing a generic form for the optimal solution, parametric in the Lagrange multipliers. He has also observed that the results can be extended to the more general minimum relative entropy problem.

We shall here extend Georgiou's approach, proving existence, uniqueness and continuity of the solution with respect to the measured data, when a generic
prior is considered. The solution can then be computed by standard numerical methods. However, the approach returns a
meaningful answer only when there is a full-rank admissible solution among the states compatible with the data. While
this appears to be a reasonable assumption as quantum full-rank states are
generic, this is no longer the case whenever the unknown state is pure or near the boundary of the physical state set.  In fact, it is
easy to picture realistic scenarios where the effect of noisy or
biased measurements might actually force the solutions to be on
the bounduary, or even outside of the admissible set \cite{Aiello-ML,ZORZI_MINIMAL_RESOURCES_2011}. In the latter case
the constrained optimization problem ``as is'' is not feasible, and one has to relax the constraints.

Our strategy to solve these issues is organized as follows: a general
setting for posing the feasibility problem and quantum minimum
relative entropy with data corresponding to linear constraints is
presented in Section \ref{Sez_quantum_state_estimation}. In Section
\ref{sec_feasibility}, we propose a way to reformulate the {\em
feasibility analysis} as a convex optimization problem. The solution
of this ancillary problem, for which we provide a numerical approach
in Appendix \ref{sec_nummethods}, also indicates an optimal way to
perturb, or {\em relax} the constraints in order to allow for
admissible solutions to the estimation problem. We also show how the
way in which the constraints are relaxed can be tailored to match
the error distribution or the level of noisy we assume on the
measurements. Once the feasibility analysis returns a positive
answer, our approach directly leads to the construction of a {\em
reduced problem} for which there exists a positive definite state
satisfying the given constraints. In Section \ref{max_ent_section},
we address the corresponding (reduced, if needed) minimum relative
entropy problem, showing it admits a unique full-rank solution. The
latter can be computed from the closed-form solution of the primal
problem, and a standard numerical algorithm to find the
corresponding Lagrange multipliers is suggested. Then, the solution
to the non-reduced, original problem is immediately obtained. Some
concluding remarks and future directions and applications are
summarized in Section \ref{conclusions}.


\section{Problem Setting}\label{Sez_quantum_state_estimation}

\subsection{Quantum States and Measurement Data}
Consider a quantum $n$-level system. Its state is described by a
density operator, namely by a positive semidefinite unit-trace
matrix \eq \rho\in\Dc_n=\Set{\rho\in\Cs^{n\times n}\;|\;
\rho=\rho^\dag\geq 0,\; \tr(\rho)=1},\eeq which plays the role of
probability distribution in the classical probability framework. Note that
a density matrix depends on $n^2-1$ real parameters.

In this work we will be concerned on the problem of reconstructing
an unknown $\rho$ from a set of repeated measurement data. This is
of course an {\em estimation} problem in the statistical language,
while in the physics community it is usually referred to as {\em
state tomography} \cite{PARIS_QSTATE_ESTIMATION_2004}.

We assume that data are provided in one of the following forms:

\subsubsection{Outcome frequencies for projective measurements} consider repeated measurements of a (Hermitian) {\em observable} \cite{sakurai}, $O=\sum_k o_{k} \Pi_{k}$, where $\{\Pi_{k}\}$ is the associated
spectral family of orthogonal projections. The spectrum $\{o_k\}$
represents the possible outcomes at each measurement, and the
frequency of the $k$-th outcome given a state $\rho$ can be
computed as $p_k=\tr(\rho \Pi_k)$. After $K$ measurements of $O,$
we assume we are provided with some experimental estimates of
$p_k,$ i.e. the experimental relative frequencies of occurences
$\hat p_k= \#(O=o_k)/K,$ with $\#(O=o_k)$ the number of
measurements that returned outcome $o_k.$

\subsubsection{Observable averages} consider a set of $n_o$ measured observables, represented by Hermitian matrices $O_i$, where now we only have access to the mean values of the outcomes, denoted by $\langle O_i\rangle$ (and with possible outcomes $o_{i,k}$), that can be theoretically computed as $\langle O_i\rangle:=\tr(\rho O_i)$ and experimentally estimated by $\langle \hat O_i\rangle = \sum_k o_{i,k}\hat p_{k}.$

\subsubsection{Outcome frequencies for general measurements} consider repeated measurements of a {\em Positive-Operator Valued Measure} (POVM), that is generalized measurements that can be used to describe indirect measurements on a system of interest \cite{nielsen-chuang}. A POVM with $M$ outcomes, say $k=1,\ldots M,$ is associated to a set of non-negative operators $\{Q_k\}_{k=1}^M$ such that $\sum_k Q_k=I,$ playing the role of resolution of the identity for  projective measurements. The probability of obtaining the $k$-th outcome can be computed by $q_k=\tr(\rho Q_k),$ and experimentally estimated by
$\hat q_k$ after $K$ repeated measurements. This case in fact includes the first one, and the generalization to multiple POVM is straightforward.


In all these scenarios, data are provided as a set of real values
representing estimates $\hat f_i$ of quantities $f_i$ (that can be
either $p_i$, $\langle O_i\rangle $ or $q_i$), each associated to the state through a linear
relation of the form $f_i=\tr(\rho Z_i),$ where $Z_i$ have the role
of $\Pi_i,O_i$ or $Q_i$ described above. Clearly
$\hat{f}_i\rightarrow f_i=\tr(Z_i\rho)$ with probability one as
$N\rightarrow \infty$. This framework is quite general, and can be
adapted to include any case if the data are given as linear
constraints. Another significant situation that fits in this
framework is when reduced states of a multipartite systems are
available as data \cite{hall-consistency,ticozzi-QLS}. Finally, by the well-known Choi-Jaimlokowskii isomorphism \cite{Choi1975}, the same setting, and methods for solution, can be adapted to include estimation of quantum channels, or quantum process tomography \cite{PARIS_QSTATE_ESTIMATION_2004,ZORZI_MINIMAL_RESOURCES_2011}.

From a theoretical viewpoint, $\rho$ can be in principle
reconstructed exactly from at least $n^2-1$ averages $f_i=\tr(\rho
Z_i)$ $i=1\ldots n^2-1$ when $Z_1\ldots Z_{n^2-1}$ are observables
which do not carry redundant information, namely they form a basis
for the space of traceless Hermitian matrices. In any practical
application, however, one has to face the following issues:
\begin{enumerate}
    \item Accurate estimates $\hat{f}_i$ of $f_i$ are only obtained by
    averaging over a large quantity of trials; Often only a small set of trials is available, and/or the data are subject to significant errors;
    \item The number of observables required for a unique reconstruction of $\rho$ grows quadratically with respect to the dimension of the quantum system, and exponentially in the number of subsystems. Typically only a small subset of these is available;
\end{enumerate}

We here analyze the estimation problem when these two aspects are
taken into account. The first one will lead us to consider the
{\em feasibility} problem, that is, if the problem admits a
physically admissible solution  for the given data. Since errors
may affect the $\hat f_i,$ the reconstructed state may not be
positive semidefinite, or a valid state that satisfies the
constraints might not even exist. The second issue generically
leads to a estimation problem where more than one state satisfy
the constraints, and thus an additional criterion has to be
introduced to arrive at a unique solution. As we said, a typical
strategy in this setting is to introduce an entropic functional,
e.g. {\em relative entropy} with respect to some reference state
representing {\em a priori} information.

\subsection{Statement of the Main Problems}

Consider the setting described above, where we want to
estimate the state of an $n$-dimensional quantum systems from the
real data $\{\hat f_i\}_{i=2}^{p}, $ experimental estimates of the
quantities $f_i=\tr(Z_i\rho),$ for the Hermitian matrices
$Z_2\ldots Z_{p},$ with $p\ll n^2-1.$ In addition to these, we introduce an
auxiliary observable $Z_1=I$ and the corresponding estimate
$\hat{f}_1=1$. In this way, we include the linear
constraint $\tr(\rho)=1$ in the constraints associated with the ``data''. We wish now to solve
the following problem.

\pb \label{prob_ricerca_rho} Given $\{Z_i\}$
and $\{\hat{f}_i\}$, $i=1\ldots p$, find: \eq \mathrm \;\;
\rho\in\Hc_n, \; \; \mathrm{such\;   that}\; \;     \rho\geq
0,\;\; \hat{f}_i=\tr(\rho Z_i),\;i=1,\ldots,p. \eeq\epb

Here,
$\Hc_n$ denotes the vector space of Hermitian matrices of
dimension equal to $n$. Notice that, if we remove the positivity constraint $\rho\geq 0,$
all other constraints are linear and identify a hyperplane in
$\Hc_n$.
To our aim it is convenient to first address a simpler problem: let \[\Sc:=\Set{\rho\in\Hc_n\;
|\;\rho\geq0,\;\; \hat{f}_i=\tr(\rho Z_i)}\] be the set of the
density matrices which solve Problem \ref{prob_ricerca_rho}.

\pb [Feasibility] \label{feasibility} Determine if $\Sc$ is not empty. \epb

When the problem is feasible, in general $\Sc$ contains more than one solution, and in principle any solution in ${\cal S}$ fits the data. We focus on choosing a solution that has minimum distance with respect to an {\em a priori} state. In the same spirit of MAXENT problem, this corresponds to give maximum priority to fitting the data, and then choosing the admissible solution that is the closest (in the relative-entropy pseudo-distance) to our {\em a priori} knowledge on the systems.
To this aim,
consider the (Umegaki's) quantum relative entropy
 between
$\rho\in\Dc_{n}$, and $\tau\in\mathrm{int}(\Dc_{n})$ \cite{NIELSEN_CHUANG_QUANTUM_STATE_ESTIMATION}:
 \eq
\label{quantum_entropy}
\Ss(\rho\|\tau)=\tr(\rho\log\rho-\rho\log\tau).\eeq  Assuming $\tau$ in the interior and with the usual convention that $0\log(0)=0$, we do not have to worry about unbounded values of $\Ss(\rho\|\tau)$.

\pb [Minimum relative entropy estimation] \label{problema_entropia} Given the
observables $Z_1\ldots Z_p$ and the corresponding estimates
$\hat{f}_1\ldots\hat{f}_p$, solve \eq \underset{\rho\geq
0}{\hbox{minimize }} \Ss(\rho\|\tau) \hbox{ subject to } \tr(\rho
Z_i)=\hat{f}_i,\; i=1\ldots p.\eeq \epb Here, $\tau$ represents the
{\em a priori} information on the considered quantum system. We set
$\sigma=I$ if no information is available. In this situation,
(\ref{quantum_entropy}) is the opposite of the quantum entropy of
$\rho,$ and we obtain a MAXENT problem. Note that, the solution to
the Problem above may be singular.

\section{Feasibility Analysis}\label{sec_feasibility}

\subsection{An auxiliary problem}

We start by addressing the {\em feasibility} problem, i.e. to
determine when $\Sc$ is not empty. In addition, whenever the problem
is not feasible, we show how to determine a suitable perturbation of
the $\{\hat{f}_i\}$ that makes our problem feasible. We will show
that the corresponding $\Sc$ only contains singular density matrices
when the constraints are relaxed.


Note that, the constraints are linear in $\hat{f}_i$ and $Z_i$,
and can be linearly combined: $\alpha\hat{f}_i+\beta
\hat{f}_k=\tr[\rho(\alpha Z_i+\beta Z_k)]$ for each
$\alpha,\beta\in\Rs\setminus \{0\}$ and $i,k=1\ldots p$. Consider
the vector space generated by the observed operators, $\Span
{Z_1\ldots Z_p}$. Thus, by applying the {\em Gram-Schmidt}
process, starting with $X_1=\frac{1}{\sqrt{n}}Z_1=\frac{1}{\sqrt
n} I$, we can compute an orthonormal basis for it: \eq X_i:=
\alpha^i_iZ_i+\sum_{l=1}^{i-1} \alpha_l^i X_l,\;\;
 i=2\ldots m\leq p.\eeq By linearity, by associating these basis elements to the estimates \eqn
\bar{f}_1&:=&\frac{1}{\sqrt n}\nn\\ \bar{f}_i&:=&
\alpha^i_i\hat{f}_i+\sum_{l=1}^{i-1} \alpha_l^i \bar{f}_l,\;\;
 i=2\ldots m,\eeqn
 we obtain a new yet {\em equivalent} set of constraints:
 \eq \label{constraint_per_rho} \bar{f}_i=\tr(\rho X_i),
\; \; i=1\ldots m.\eeq Note that, $I\in\Span{X_1\ldots X_m}$.

Let $Y_1\ldots Y_{n^2-m}$ be an orthonormal completion of
$X_1\ldots X_m$ to a basis of ${\cal H}_n$. Accordingly, all the
Hermitian matrices, and in particular density operators, can be
expressed as \eq \label{parametr_rho}\rho=\sum_{i=1}^m \alpha_i
X_i+\sum_{i=1}^{n^2-m} \beta_i Y_i,\eeq  with $\tr(\rho
X_i)=\alpha_i$. In particular, all the Hermitian matrices
satisfying the linear constraints in (\ref{constraint_per_rho})
depend on $n^2-m$ parameters $\beta=[\beta_1\ldots
\beta_{n^2-m}]$: \eq \rho=\tilde\rho_0+\sum_{i=1}^{n^2-m} \beta_i
Y_i\eeq where we have defined the (not necessarily positive)
pseudo-state associated to the constraints: \eq\label{rho_0}
\tilde\rho_0=\sum_{i=1}^m \bar{f}_i X_i.\eeq

In the light of this observation, the feasibility problem consists
in checking if there exists at least one vector
$\beta\in\Rs^{n^2-m}$ such that $\rho\geq 0$. To this aim, we
introduce an auxiliary problem. Intuitively, the idea is the following: given
any Hermitian matrix $\tilde\rho_0,$ there always exists a real $\mu$
such that $\tilde\rho_0+\mu I$ is positive definite. More
precisely, if $\tilde\rho_0$ is not positive definite already, it
is easy to see that such a $\mu$ will need to be positive. On the other hand, if $\tilde\rho_0$ is already positive, the perturbed matrix
remains positive semi-definite for some small, negative $\mu$.
Studying the minimal $\mu$ that correspond to a positive
semidefinite matrix offers us a way to understand whether our
constraints allow for physically admissible solutions.

Let us formalize these idea: we define $c:=\left[%
\begin{array}{cccc}
  0 & \ldots & 0 & 1 \\
\end{array}%
\right]^T\in\Rs^{n^2-m+1}$, and
\[H(\underline{v}):=\tilde\rho_0+\sum_{i=1}^{n^2-m} v_i
Y_i+v_{n^2-m+1}X_1\] with $\underline{v}=\left[%
\begin{array}{ccc}
  v_1 &  \ldots & v_{n^2-m+1} \\
\end{array}%
\right]$ and we consider the following minimum eigenvalue problem.
\pb\label{min_eigen_pb} Given $\tilde\rho_0$ as in (\ref{rho_0})
and $Y_1\ldots, Y_{n^2-m}$ an orthonormal completion of
$\Span{X_1\ldots X_m}$, solve \eq \mathrm{minimize }\; c^T
\underline{v} \;\; \; \mathrm{subject \; to}\; \; \;
\underline{v}\in\Ic:=\Set{\underline{v}\; |\; H(\underline{v})\geq
0}.\eeq \epb

\lem Problem \ref{min_eigen_pb} always admits solution.\elem
\proof First of all, notice that Problem \ref{min_eigen_pb} is a
convex optimization problem, and the objective function $c^T
\underline{v}$ is linear and continuous over the set $\Ic$. Then,
the proof is divided in three steps.\\ {\em Step 1:}  We show that
$c^T \underline{v}=v_{n^2-m+1}$ is {\em bounded from below} on
$\Ic$: since ${X_1,\ldots, X_m,Y_1,\ldots,Y_{n^2-m}}$ forms an
orthonormal basis and $I\in\Span{X_1,\ldots ,X_m}$, the matrices
$\{Y_i\}$ are traceless. Thus,
\begin{eqnarray*}\tr[H(\underline{v})]&=&\tr[\tilde\rho_0+\sum_{i=1}^{n^2-m}v_i
Y_i+v_{n^2-m+1}X_1]\\&&=\tr(\tilde\rho_0)+\sqrt nv_{n^2-m+1}=1+\sqrt n
v_{n^2-m+1}\end{eqnarray*} and $\tr[H(\underline{v})]\geq 0$ for each
$\underline{v}\in \Ic$. Hence,
$c^T\underline{v}=v_{n^2-m+1}\geq -\frac{1}{\sqrt n}$ for each $\underline{v}\in\Ic$.\\
{\em Step 2:}  Let us consider $\underline{v}_0=\left[%
\begin{array}{cccc}
  0 & \ldots & 0 & \sqrt n(-\lambda_{\mathrm{min}}(\tilde\rho_0)+1) \\
\end{array}%
\right]\in\Ic$ where $\lambda_{\mathrm{min}}(\tilde\rho_0)$
denotes the minimum eigenvalue of $\tilde\rho_0$. Accordingly,
$c^T\underline{v}_0=\sqrt
n(-\lambda_{\mathrm{min}}(\tilde\rho_0)+1)$ and Problem
\ref{min_eigen_pb} is equivalent to minimize $c^T \underline{v}$
over the closed sublevel set $\Ic_0=\{\underline{v}\;|\; \;
H(\underline{v})\geq 0,\;\; -\frac{1}{\sqrt n}\leq v_{n^2-m+1}\leq
\sqrt n (-\lambda_{\mathrm{min}}(\tilde\rho_0)+1)\}\subset\Ic$.
We want to show that $\Ic_0$ is {\em bounded} and accordingly {\em
compact} (recall that we are working in a finite dimensional space).
This can done by proving that a sequence
$\Set{\underline{v}^k}_{k\geq 0}$ such that
$\|\underline{v}^k\|\rightarrow \infty$ cannot belong to $\Ic_0$. It
is therefore sufficient to show that the minimum eigenvalue of the
associated Hermitian matrix $H(\underline{v}^k)$ tends  to $-\infty$
as $\|\underline{v}^k\|\rightarrow \infty$ with $v_{m^2-n+1}$
bounded. Note that the affine map $\underline{v}\mapsto
H(\underline{v})$ is injective, since $Y_1\ldots Y_{n^2-m},X_1$ are
linearly independent. Accordingly
$\|H(\underline{v}^k)\|\rightarrow\infty$ as
$\|\underline{v}^k\|\rightarrow\infty$. Since $H(\underline{v}^k)$
is an Hermitian matrix, $H(\underline{v}^k)$ has an eigenvalue
$\eta_k$ such that $|\eta_k|\rightarrow\infty$ as
$\|\underline{v}^k\|\rightarrow \infty$. By construction
$\tr[H(\underline{v}^k)]=1+\sqrt n v_{n^2-m+1}^k$ and
$v_{n^2-m+1}^k$ is bounded in $\Ic_0$. Thus
$\tr[H(\underline{v}^k)]<\infty$, namely the sum of its eigenvalues
is always bounded. Thus, there exists an eigenvalue of
$H(\underline{v}^k)$ which approaches $-\infty$ as $k\rightarrow
\infty$. So, $\Ic_0$ is bounded.\\ {\em Step 3:} Since
$c^T\underline{v}$ is continuous over the compact set $\Ic_0$, by
Weiestrass' theorem we conclude that
$c^T \underline{v}$ admits a minimum point over $\Ic_0$.\qed\\

We need to take into account that the vector which minimizes
$c^T\underline{v}$ over $\Ic$ may not be in general unique. However, to our aim we are more interested in the sign of the minimum. 
\prop \label{propsing} Let $\mu=\Min{\underline{v}\in\Ic}
c^T\underline{v}$. Then, the following facts
hold:\begin{enumerate}
    \item If $\mu>0$, then Problem \ref{prob_ricerca_rho} is not
    feasible
    \item If $\mu <0$, then Problem \ref{prob_ricerca_rho} is
    feasible and there exists at least one positive definite matrix satisfying
    constraints in (\ref{constraint_per_rho})
    \item If $\mu=0$, then Problem \ref{prob_ricerca_rho} is
    feasible and all the matrices satisfying constraints in (\ref{constraint_per_rho}) are
    singular.
\end{enumerate}   \eprop
\proof Note that $G(v_1,\ldots,
v_{n^2-m}):=\tilde\rho_0+\sum_{i=1}^{n^2-m} v_iY_i$ represents
the parametric family of Hermitian matrices (not necessary
positive semidefinite) satisfying constraints in
(\ref{constraint_per_rho}). Define $\varepsilon:=-v_{n^2-m+1}$,
thus Problem \ref{min_eigen_pb} can be rewritten in the following
way:\eq \label{pb_min_eigen_val_riscritto_in_eps}\hbox{maximize}
\; \varepsilon \; \; \hbox{subject to} \; \; G(v_1,\ldots,
v_{n^2-m})\geq \frac{\varepsilon}{\sqrt n} I.\eeq Let
$\varepsilon^\circ=-\mu$ be the solution of the above problem. If
$\varepsilon^\circ<0$, the parametric family $G(v_1,\ldots,$ $
v_{n^2-m})$ does not contains positive semidefinite matrices,
accordingly Problem \ref{prob_ricerca_rho} does not admit
solution. If $\varepsilon^\circ\geq 0$, the parametric family
$G(v_1,\ldots, v_{n^2-m})$ contains at least one positive
semidefinite matrix and Problem \ref{prob_ricerca_rho} admits
solution. Moreover if $\varepsilon^\circ>0$, the parametric family
contains at least one matrix $\rho\geq
\frac{\varepsilon^\circ}{\sqrt n} I>0$ which is positive definite.
On the contrary, for $\varepsilon^\circ=0$ there only exist
positive semidefinite matrices which are singular.
\qed\\

An effective numerical approach for the solution to the problem is
described in Appendix \ref{sec_nummethods}.

In the light of the previous result, if  $\mu<0$ Problem
\ref{prob_ricerca_rho} is feasible and $\Sc$ contains at least one
positive definite solution. As we will see in Section
\ref{max_ent_section}, this condition ensures that a minimum relative entropy criterion will lead to an admissible
solution in $\Sc$. The remaining cases need to be studied more
carefully. We start by showing how to make Problem
\ref{prob_ricerca_rho} feasible when it is not be so for the
given constraints. It turns out that a minimally relaxed problem is
feasible and $\Sc$ only contains singular density matrices. Next,
we deal with the case in which Problem \ref{prob_ricerca_rho} is
feasible and all its solutions are singular, showing how they all
share a minimal kernel and how to construct a reduced problem with
a full-rank solution for which the minimum relative entropy
methods work.

\subsection{Forcing the feasibility condition (case
$\mu>0$)}\label{sec_feasibility_forcing}

The parameter $\mu$ given by the auxiliary problem described above
reveals if the original problem is feasible, but also suggests an
``optimal'' way to relax unfeasible constraints so that they make
Problem \ref{prob_ricerca_rho} feasible. In fact, from the
definition of $\mu$, we know that there exist $v_1\ldots
v_{n^2-m}\in\Rs$ such that
\[\tilde\rho_\mu:=\tilde\rho_0+\sum_{i=1}^{n^2-m} v_iY_i+\mu X_1\geq 0,\]
and:
\[\tr(\tilde\rho_\mu)=1+\sqrt n \mu .\]
From this positive operator, in order to obtain a density operator, we only need to normalize the trace by defining:
\begin{eqnarray}\label{relax}\rho&:=&\frac{1}{1+\sqrt n\mu}\tilde\rho_\mu\\&=&\frac{1}{\sqrt n}X_1+\sum_{i=2}^m \frac{\bar{f}_i}{1+\sqrt n\mu} X_i+\sum_{i=1}^{n^2-m} \frac{v_i}{1+\sqrt n \mu}Y_i.\nonumber \end{eqnarray}
This implies that the original problem can be made feasible by
uniformly, ``isotropically'' {\em contracting} the data
$\{\bar{f}_i\}$ of a factor ${1}/({1+\sqrt n\mu})$ and, in light
of the fact that $\mu$ is a solution to Problem
\ref{min_eigen_pb}, that this is the {\em minimum} amount of
contraction that makes Problem \ref{prob_ricerca_rho} feasible.
Moreover, the corresponding set $\Sc$ only contains singular
solutions.

However, the entries in the data set $\{\bar{f}_i\}$ may differ in
their reliability, and one would like to be able to relax the
corresponding constraints accordingly. This is complicated by the
fact that the original $\{Z_i\}$ may not be orthogonal, and the
data we are contracting are in fact the linearly transformed
output of the Gram-Schmidt orthonormalization described above.

This weighed relaxation can be realized as follows: consider the
initial setting of Section \ref{Sez_quantum_state_estimation},
where we have $p$ observables $Z_1\ldots Z_p$ (not necessarily
orthonormal), with $Z_1=I$ and $\hat{f}_1=1$. Define the {\em
reliability indexes} $0<d_2\ldots d_p\leq1$ associated to each
observable $Z_2\ldots Z_p$. More precisely, the more $\hat{f}_i$
is reliable, the closer to one $d_i$ is. This information can be
extracted, for example, from an error analysis on the measurement
procedures, with $d_i$ associated to the normalized reciprocal of
the variances.

When we obtain the orthonormal generators $X_1 \ldots X_m$, the {\em
Gram-Schmidt} process induces a linear transformation on the
original estimates $\hat{f}_1,\ldots, \hat{f}_p$: \eq \left[
                         \begin{array}{c}
                           \bar{f}_1 \\
                           \vdots \\
                           \bar{f}_m \\
                         \end{array}
                       \right] = T\left[
                                  \begin{array}{c}
                                    \hat{f}_1 \\
                                    \vdots \\
                                    \hat{f}_p \\
                                  \end{array}
                                \right]\eeq
where $T=\left[
           \begin{array}{cc}
             \frac{1}{\sqrt n} & 0 \\
             T_1 & T_2 \\
           \end{array}
         \right]\in\Rs^{m\times p}
$.

In order to modify the data $\{\hat f_i\}$ according to their
reliability indexes, we define the new set of data: \eq \left[
                         \begin{array}{c}
                           \hat{f}_{1}' \\
                           \vdots \\
                           \hat{f}_{m}' \\
                         \end{array}
                       \right] = T \left[
                                  \begin{array}{c}
                                   \hat{f}_1 \\
                                     k{d_2}\hat{f}_2 \\
                                    \vdots \\
                                     k{d_p} \hat{f}_p \\
                                  \end{array}
                                \right].\eeq
where $k>\max\{d_i^{-1}\}.$ In this way
$\hat{f}_2,\ldots,\hat{f}_p$ are amplified of a factor
$k{d_i}> 1$ according their reliability indexes. This
will allow for the maximum contraction to be applied to the most
noisy estimates.

In order to compute the {\em minimum} $\mu$ that makes the original
problem feasible perturbing the data consistently with their
reliability indexes, we can solve Problem \ref{min_eigen_pb} with
respect to the new pseudo-state: \eq\label{rho_0prime}
\tilde\rho_0'=\sum_{i=1}^m \hat{f}'_i X_i.\eeq
 It is easy to see that if the original problem was unfeasible, this modified problem is unfeasible as well for $k$ large enough, since all the $\hat f_i$ corresponding to traceless operators are multiplied for a factor $K{d_i}>> 1$.
Let $\mu^\prime>0$ be the parameter given by the auxiliary problem
when $\tilde\rho_0^\prime$ is considered. By the results of the
previous subsection and \eqref{relax} above, we consider the
perturbed
constraints \eqn \bar{f}_1&=& \frac{1}{\sqrt n}\nn\\
\bar{f}_{i}&=& \frac{1}{1+\sqrt n \mu'} \hat{f}_{m}',\;\;
i=2\ldots m.\eeqn Thus, the corresponding Problem
\ref{prob_ricerca_rho} is feasible and $\Sc$ only contains
singular solutions.

\subsection{Case $\mu=0$}

In the limit case $\mu=0$, not only all solutions are singular,
but they share a key property. \prop Assume that, with the
definition above, $\mu=0.$ Then there exists {a kernel ${\cal
K}$ which is common} for all $\rho\in {\cal S}.$ \eprop \proof Let
us assume $\mu=0,$ accordingly Problem 1 does only admit singular
solutions, with $\mathrm{dim}\ker(\rho)>0$ $\forall \; \rho
\in\Sc$. Pick a solution $\rho^\circ\in\Sc$ with kernel of minimal
dimension. Suppose by contradiction that there exists
$\bar{\rho}\in\Sc$ such that $\ker (\rho^\circ)\nsubseteq
\ker(\bar{\rho})$. Taking into account $p\in(0,1)$, we define
$\rho:=p\rho^\circ +(1-p)\bar{\rho}\in\Sc$. Accordingly
$\mathrm{dim}\ker(\rho)<\mathrm{dim}\ker(\rho^\circ)$ which is a
contradiction, since $\rho^\circ$ has kernel with minimal
dimension on $\Sc$. We conclude that
$\Kc=\ker(\rho^\circ)\subseteq \ker (\rho)$ $\forall
\;\rho\in\Sc$. \qed

This directly implies the following block-form for all the solutions
to Problem 1. \cor\label{cor_param_set_soluzioni_singolari} Let
$\rho^\circ\in\Sc$
be a solution with minimal kernel ${\cal K}$ and consider its block-diagonal form \[\rho^\circ=U\left[%
\begin{array}{cc}
  \rho_1^\circ & 0 \\
  0 & 0 \\
\end{array}%
\right]U^\dag,\]
where $U$ is a unitary change of basis consistent with the Hilbert space decomposition $\Hc={\cal K}^\perp\oplus{\cal K}$ so that  $\rho_1^\circ>0$. Then, the set of all the
solutions of Problem
\ref{prob_ricerca_rho} is \eq \Sc=\Set{\rho=U\left[%
\begin{array}{cc}
  \rho_1& 0 \\
  0 & 0 \\
\end{array}%
\right]U^\dag\;|\; \rho_1\geq 0,\; \bar f_i=\tr(\rho X_i)}.\eeq
\ecor

As consequence of Corollary \ref{cor_param_set_soluzioni_singolari},
we can focus on a reduced version of Problem \ref{prob_ricerca_rho},
by considering optimization only on the support of $\rho^\circ,$ for
which the minimum relative entropy is applicable since
$\rho_1^\circ$ is positive definite, see Section
\ref{max_ent_section}.

\section{State Estimation with Minimum Relative Entropy Criterion} \label{max_ent_section}

\subsection{Reduced Problem}
In the previous part of the paper we showed how to check the
feasibility of Problem \ref{prob_ricerca_rho} given the
constraints associated to the data and, if needed, how to relax
the constraints in such a way that the corresponding Problem
\ref{prob_ricerca_rho} is feasible. In general, however, the set
of solutions $\Sc$ is not constituted by only one element. In this
section, we show how to choose, and then compute, a solution in
$\Sc$ according the {\em minimum quantum relative entropy}
criterion.

Given the results of the previous sections, we can assume that
either Problem \ref{prob_ricerca_rho} admits at least one (strictly)
positive definite solution, or we can resort to a reduced problem
for which a full rank solution exists. In fact, if $\Sc$ only
contains singular matrices (case $\mu=0,$ or after relaxation of the
constraints), by Corollary
\ref{cor_param_set_soluzioni_singolari} we have that the set of solution is \eq \Sc=\Set{\rho=U\left[%
\begin{array}{cc}
  \rho_1& 0 \\
  0 & 0 \\
\end{array}%
\right]U^\dag\;|\; \rho_1\geq 0,\; \bar f_i=\tr(\rho X_i)}\eeq for
some unitary change of basis $U$ consistent with the Hilbert space
partition $\Hc={\cal K}^\perp\oplus{\cal K}$. Accordingly for each
$\rho\in\Sc$, constraints in (\ref{constraint_per_rho}) can be
rewritten in the following way
\eq \bar{f}_i=\tr(\rho X_i)=\tr\left(U \left[%
\begin{array}{cc}
  \rho_1& 0 \\
  0 & 0 \\
\end{array}%
\right]U^\dag X_i\right)=\tr(\rho_1 \bar{X}_i)\eeq where
$\bar{X}_i:=
\left[%
\begin{array}{cc}
  I & 0 \\
\end{array}%
\right]U^\dag X_1U\left[%
\begin{array}{c}
  I \\
  0 \\
\end{array}%
\right]\in\Hc_{n_1}$ with $n_1<n$. Accordingly Problem
\ref{prob_ricerca_rho} is equivalent to the corresponding reduced
problem with $\bar{X}_1\ldots \bar{X}_m$ and $\bar{f}_1\ldots
\bar{f}_m$. The corresponding set of solutions is \eq
\Sc_1=\Set{\rho_1\in\Hc_{n_1}\;|\; \rho_1\geq 0,\;
\bar{f}_i=\tr(\rho_1\bar{X}_i)}\eeq which contains the positive
definite solution $\rho^\circ_1$. Once chosen a solution
$\hat{\rho}_1\in\Sc_1$, the
corresponding original solution is \eq\hat{\rho}=U\left[%
\begin{array}{cc}
  \hat{\rho}_1 & 0 \\
  0 & 0 \\
\end{array}%
\right]U^\dag.\eeq In the effort of keeping a simple notation, we
will not distinguish between the reduced and the full problem in the
following discussion, therefore using $\rho$ for either the full or
the reduced state, $\{X_i\}$ for either the full or reduced
observable, and $n$ for the dimension of the full Hilbert space or
the reduced one as needed. We can consider the following simpler
problem, restricted to strictly positive matrices:
\pb \label{problema_entropia1} Given the observables $ X_1\ldots
 X_m$ and the corresponding estimates $\bar{f}_1\ldots\bar{f}_m$,
solve \eq \underset{\rho> 0}{\hbox{minimize }} \Ss(\rho\|\tau)
\hbox{ subject to } \tr(\rho \bar X_i)=\bar{f}_i,\; i=1\ldots m.\eeq
\epb


\subsection{Lagrangian and Form of the Full-Rank Solution}

Now we are ready to derive a solution method for problem the entropic criterion.
Consider the linear operator associated to the above constraints:
\eqn L&:& \Hc_n \rightarrow \Rs^{m}\nn\\ && \rho \mapsto \left[
                                             \begin{array}{c}
                                               \tr(\rho X_1) \\
                                               \vdots \\
                                               \tr(\rho X_m) \\
                                             \end{array}
                                           \right].
\eeqn Given $\lambda=\left[
                       \begin{array}{ccc}
                         \lambda_1 & \ldots & \lambda_m \\
                       \end{array}
                     \right]^T\in\Rs^{m}
$ and $\rho\in\Hc_n$, \eq
\Sp{L(\rho)}{\lambda}=
\sum_{i=1}^m \lambda_i\tr(\rho X_i)= \tr(\rho \sum_{i=1}^m
\lambda_i X_i)=\Sp{\rho}{L^*(\lambda)}\eeq where \eqn L^*&:&
\Rs^{m}\rightarrow \Hc_n\nn\\ && \lambda \mapsto \sum_{i=1}^m
\lambda_i X_i\eeqn is the adjoint operator of $L$.
Define $\bar{f}=\left[%
\begin{array}{ccc}
  \bar{f}_1& \ldots & \bar{f}_m \\
\end{array}%
\right]^T$. Since Problem \ref{problema_entropia1} is a constrained
convex optimization problem, we take into account its Lagrangian
\eqn
\Lc(\rho,\lambda)&=&\tr(\rho\log\rho-\rho\log\tau)-\Sp{\lambda}{\bar{f}-L(\rho)}\nn\\
&=&
\tr(\rho\log\rho-\rho\log\tau)+\Sp{L^*(\lambda)}{\rho}-\Sp{\lambda}{\bar{f}}\nn\\
&=&
\tr[\rho(\log\rho-\log\tau+L^*(\lambda))]-\Sp{\lambda}{\bar{f}}\eeqn
where $\lambda\in\Rs^{m}$ is the Lagrange multiplier. Note that
$\Lc(\cdot,\lambda)$ is strictly convex over $\Hc_{n,+}$ where
$\Hc_{n,+}$ denotes the cone of the positive definite matrices.
Thus, its minimum point is given by annihilating its first
variation \eqn
\delta\Lc(\rho,\lambda;\delta\rho)&=&\tr[(\log\rho+I-\log\tau+L^*(\lambda))\delta\rho]\eeqn
for each direction $\delta\rho\in\Hc_n$. Accordingly, the unique
minimum point for $\Lc(\cdot,\lambda)$ is \eq \label{forma_ottimo}
\rho(\lambda)=\e{\log\tau-I-L^*(\lambda)}\eeq and \eq
\label{inequality_lagrangiana} \Lc(\rho(\lambda),\lambda)\leq
\Lc(\bar\rho,\lambda),\; \forall \;\bar\rho \in\Hc_{n,+}.\eeq If
there exists $\lambda^\circ$ such that
$\rho(\lambda^\circ)\in\Sc$, i.e. $\bar f=L(\rho(\lambda^\circ))$,
then (\ref{inequality_lagrangiana}) implies \eq
\Ss(\rho(\lambda^\circ)\|\tau)\leq \Ss(\bar\rho\|\tau),\;\forall
\; \bar\rho\in\Sc.\eeq Thus, if we are able to find
$\lambda^\circ\in\Rs^m$ such that $\bar
f-L(\rho(\lambda^\circ))$=0, then $\rho(\lambda^\circ)$ is the
unique solution to Problem \ref{problema_entropia1}. This issue is
solved by considering the dual problem wherein $\lambda^\circ$ (if
there exists) maximizes the following functional over $\Rs^m$
 \eqn \label{funzionale_duale_camb_segno}
\Inf{\rho\in\Hc_{n,+}}
\Lc(\rho,\lambda)&=&\Lc(\rho(\lambda),\lambda)\nonumber\\&=&-\tr(\e{\log\tau-I
-L^*(\lambda)})-\Sp{\lambda}{ \bar{f}}.\eeqn The existence of such
a $\lambda^\circ$ is proved in Section \ref{problema duale}.
Moreover, we suggest how to efficiently compute it.

\rem When $\mu=0$, $\Sc$ only contains singular matrices. Instead of
considering the reduced problem as we did, one could try to consider
Problem \ref{problema_entropia} with relaxed constraint $\rho\geq
0$. In this situation the {\em Slater's condition}
\cite[5.2.3]{BOYD_CONVEX_OPTIMIZATION}, however, does not hold
because $\Sc$ does not contain positive definite matrices. Hence, we
cannot conclude that $\rho(\lambda^\circ)$ is the desired solution
of the primal problem. Moreover,  for any $\lambda,$ would be
$\rho(\lambda^\circ)>0$. This means that $\rho(\lambda^\circ)$ is
not the solution to Problem \ref{problema_entropia}. \erem

\subsection{Dual Problem: Existence and Uniqueness of the Solution} \label{problema duale}
The dual problem consists in maximizing
(\ref{funzionale_duale_camb_segno}) over $\Rs^m$ which is equivalent
to minimize \eq J(\lambda)=\tr(\e{\log\tau-I
-L^*(\lambda)})+\Sp{\lambda}{\bar{f}}.\eeq This functional will be
referred to as {\em dual function} throughout this Section. Before
to prove the existence of $\lambda^\circ$ which minimizes $J$ we
need to introduce the following technical results.

First of all, note that $\Sp{\lambda^\bot}{\bar{f}}=0$ for each
$\lambda^\bot\in[\Range L]^\bot$. In fact, if $\lambda^\bot
\in[\Range L]^\bot=\ker L^*$, then $L^*(\lambda^\bot)=0$. Since
$\Sc\cap \Hc_{n,+}\neq \emptyset$, there exists $\rho_f\in\Hc_{n,+}$
such that $\bar{f}=L(\rho_f)$. Thus,
\eq\Sp{\lambda^\bot}{\bar{f}}=\Sp{\lambda^\bot}{L(\rho_f)}=\Sp{L^*(\lambda^\bot)}{\rho_f}=\tr(L^*(\lambda^\bot)\rho_f)=0.\eeq
We conclude that $\lambda^\bot$ does not affect $J$, i.e.
 \eq J(\lambda+\lambda^\bot)=J(\lambda),\;\; \; \forall \;
\lambda^\bot\in[\Range L]^\bot. \eeq We may therefore restrict the
search of the minimum point for $J$ over $\Range L$.  \prop $J$ is
strictly convex over $\Range L$.\eprop \proof Since $J$ is the
opposite of $\Lc(\rho(\lambda),\lambda)$, it is convex over
$\Rs^{m}$.
 The first and the second variation of
$J(\lambda)$ in direction $\delta\lambda\in\Rs^{m}$ are: \eqn\delta
J(\lambda;\delta\lambda)&=&-\tr\int_0^1 \LARGE{(}\e{(1-t)(\log\tau-I
-L^*(\lambda))} L^*(\delta\lambda)\\
&&\hspace{11mm}\cdot\;\e{t(\log\tau-I
-L^*(\lambda))}\LARGE)\de t+\Sp{\delta\lambda}{\bar{f}}\nn\\
&=&-\tr\int_0^1 \e{\log\tau-I -L^*(\lambda) }\de t L^*(
\delta\lambda)
+\Sp{\delta\lambda}{\bar{f}}\nn\\&=&-\tr(\e{\log\tau-I
-L^*(\lambda)}L^*(\delta\lambda)) +\Sp{\delta\lambda}{\bar{f}}\eeqn
\eqn\delta^2 J(\lambda;\delta\lambda)&=&\tr [\int_0^1
\e{(1-t)(\log\tau-I -L^*(\lambda))} L^*(\delta\lambda)\nonumber\\
&&\hspace{9mm}\cdot\;\e{t(\log\tau-I
-L^*(\lambda))}L^*(\delta\lambda)\de t].\eeqn Here, we exploited
the expression for the differential of the matrix exponential (see
\cite[Appendix IA]{RELATIVE_ENTROPY_GEORGIOU_2006}). Define
$Q_t=\e{t(\log\tau-I -L^*(\lambda))}$ which is positive definite
for each $t\in\Rs$. Thus,
\eqn\delta^2J(\lambda;\delta\lambda)&=&\int_0^1 \tr (Q_{1-t}
L^*(\delta\lambda) Q_tL^*(\delta \lambda))\de t\\& =&\int_0^1 \tr
(Q_t^{\frac{1}{2}}L^*(\delta\lambda) Q_{1-t} L^*(\delta\lambda)
Q_t^{\frac{1}{2}})\de t\geq 0. \nn\eeqn Assume now that
$\delta\lambda\in \Range L$. If
$\delta^2J(\lambda;\delta\lambda)=0$, then $\tr
(Q_t^{\frac{1}{2}}L^*(\delta\lambda) Q_{1-t}
L^*(\delta\lambda)Q_t^{\frac{1}{2}})=0$. Since $Q_t>0$ for each
$t\in\Rs$, it follows that $L^*(\delta\lambda)=0$. Since
$\delta\lambda\in\Range L$, we get $\delta \lambda=0$. We conclude
that $\delta^2J(\lambda;\delta\lambda)>0$, for each
$\delta\lambda\neq 0$, i.e. the statement holds.\qed\\ In the
light of the previous result, the dual problem admits at most one
solution, say $\lambda^\circ$, over $\Range L$. If such a
$\lambda^\circ$ does exist, then $\delta
J(\lambda;\delta\lambda)=0$ $\forall \; \delta\lambda\in\Range L$
which is equivalent to $-L(\e{\log \tau
-I-L^*(\lambda^\circ)})+\bar{f}=0$. It means that
$\rho(\lambda^\circ)$ satisfies constraints in
(\ref{constraint_per_rho}) and it is therefore the unique solution
to Problem \ref{problema_entropia1}. It remains to show that such
a $\lambda^\circ$ does exist.  \prop $J$ admits a minimum point
over $\Range L$.\eprop \proof  We have to show that the continuous
function $J$ takes minimum value over $\Range L$. Observing that
$J(0)=\frac{1}{e}\tr(\tau)$, we can restrict our search over the
closed set \[\Mc:=\Set{\lambda\in\Rs^{m}\;|\; J(\lambda)\leq
J(0)}\cap \Range L.\] We shall show that $\Mc$ is bounded. To this
aim, consider a sequence $\Set{\lambda_i}_{i\in\Ns}$,
$\lambda_i\in\Range L$ such that $\|\lambda_i\|\rightarrow
\infty$. It is therefore sufficient to show that
$J(\lambda_i)\rightarrow \infty$, as $i\rightarrow \infty$. First
of all note that the minimum singular value $\alpha$ of $L^*$
restricted to $\Range L=[\ker L^*]^\bot$ is strictly positive,
accordingly \eq \|L^*(\lambda_i)\|\geq \alpha
\|\lambda_i\|\rightarrow \infty.\eeq This means that
$L^*(\lambda_i)$, which is an Hermitian matrix, has at least one
eigenvalue $\beta_i$ such that $|\beta_i|$ approach infinity. If
$\beta_i\rightarrow -\infty$, then the first term of $J$ tends to
infinity and dominates the second one, accordingly
$J(\lambda_i)\rightarrow \infty$. In the remaining possible case
no eigenvalue of $L(\lambda_i)$ approaches $-\infty$ and
$\beta_i\rightarrow \infty$. Thus, $L^*(\lambda_i)\geq M I$ where
$M\in\Rs$ is a finite constant and the first term of $J$ takes a
finite value. Since $\Sc\cap\Hc_{n,+}$ is not empty, there exists
$\rho_f\in \Hc_{n,+}$ such that $\bar{f}=L(\rho_f)$ and
\eq\Sp{\lambda_i}{\bar{f}}=\Sp{L^*(\lambda_i)}{\rho_f}=\tr(\rho_f^{\frac{1}{2}}L^*(\lambda_i)\rho_f^{\frac{1}{2}})\geq
M,\eeq where we exploited the fact that $\tr(\rho_f)=1$. This
means that $\Sp{\lambda_i}{\bar{f}}$ cannot approach $-\infty$.
Finally, $\|\rho_f^{\frac{1}{2}}L^*(\lambda_i)\rho_f^{\frac{1}{2}}
\|\rightarrow \infty$, because $\rho_f>0$. It follows that
$\rho_f^{\frac{1}{2}}L^*(\lambda_i)\rho_f^{\frac{1}{2}}$, and
hence $L^*(\Lambda_i)$, have at least one eigenvalue tending to
$\infty$. Accordingly $J(\lambda_i)\rightarrow \infty$ as
$i\rightarrow \infty$. We conclude that $\Mc$ is bounded and
accordingly
compact. By Weierstrass' theorem,  $J$ admits minimum point over $\Mc$.\qed\\
Finally, $\lambda^\circ$ may be computed by {\em e.g.} employing a {\em
Newton algorithm with backtracking}, see Section VI in
\cite{ME_ENHANCEMENT_FERRANTE_2012}, which globally converges.

\section{Conclusions}\label{conclusions}

The proposed set of analytic results and algorithms provides a
general method to find the {\em exact minimum relative entropy
estimate of a quantum state under general assumptions}, that is,
without requiring the constraints to include a full-rank solution
to begin with. A general solution to the problem was missing for
quantum estimation, and our feasibility analysis is of interest
for the classical case as well. Summarizing, we have proposed: (i)
A numerical method to decide the feasibility of Problem
\ref{prob_ricerca_rho}, which is of interest on his own, and
compute the minimum necessary relaxation of the constraints
whenever necessary to obtain at least one solution; (ii) As a side
product, we are able to determine whether there is at least a
full-rank solution. When this is not the case, we proved there
exists, and devised a way to determine, the maximal common kernel
of all the solution; (iii) We extended Georgiou's approach to
maximum entropy estimation to our setting, and analyzed in depth
both the primal and the dual optimization problems. The general
form of a full-rank solution of the reduced problem, namely the
one we obtain by removing the common kernel, is given explicitly
and depends on the solution of the dual problem. The latter is
proven to be a convex optimization problem with a unique solution,
$\lambda^\circ$, which can be obtained by standard numerical
algorithms.

It is also possible to show, $\lambda^\circ$ is continuous with respect to the data set $\bar f$ (see Appendix \ref{continuity_lambda0}). Since, in view of
(\ref{forma_ottimo}), $\rho(\lambda^\circ)$ is continuous with
respect to $\lambda^\circ$, we can conclude that the solution to Problem
\ref{problema_entropia1} is continuous with respect to the data $\bar
f_1 \ldots \bar f_m$. This is of course a desirable property, and ensures that for a increasing accuracy of the estimates, the computed solution will converge to the actual state.

Possible extensions of the present framework include applications to state reconstruction from local marginals and its connection to entanglement generation \cite{hall-consistency,ticozzi-QLS}, as well as comparison with the existing approximate results \cite{PARIS_MIN_KL_2007} in physically meaningful situation.
Lastly, the advantage offered by the introduction of an {\em a priori} state in the estimation problem can be exploited to devise recursive algorithms, that update existing estimate in an optimal way relying only on partial data.

\appendix

\subsection{Numerical solution for the feasibility problem}\label{sec_nummethods}
We propose a Newton-type algorithm with logarithmic barrier for
numerically find one solution $\underline{v}^\circ$ to Problem
\ref{min_eigen_pb}. Using the same approach of \cite[Section
4]{ZORZI_MINIMAL_RESOURCES_2011}, we resort to the approximate
problem {\mz \eq
\Min{\underline{v}\in\mathrm{int}(\Ic)}g_q(\underline{v})\label{approxprob}\eeq
where $q>0$, and \eq
g_q(\underline{v}):=qc^T\underline{v}-\log\det(H(\underline{v})).\eeq
Recall that we defined
\[\Ic=\{\underline{v}\;|\; \;
H(\underline{v})\geq 0\},\] and notice that $g_q$ is continuous
and strictly convex over the set $\mathrm{int}(\Ic)$. Moreover
$\lim_{\underline{v}\rightarrow
\partial \Ic} g_q(\underline{v})=+\infty$ and the component $v_{n^2-m+1}$
can be restricted to belong to a closed and bounded interval.
Accordingly the approximated problem admits a unique solution,
denoted by $\hat{\underline{v}}^q$, which is numerically computed
by employing the Newton algorithm with backtracking,
\cite[9.5]{BOYD_CONVEX_OPTIMIZATION}. Here, we can
choose as initial guess the vector $\hat{\underline{v}}^q_0=\left[%
\begin{array}{cccc}
  0 & \ldots & 0 &  \sqrt n(-\lambda_{\mathrm{min}}(\tilde\rho_0)+1 \\
\end{array}%
\right]^T\in\mathrm{int}(\Ic)$. Concerning the $l$-th Newton step,
we have to solve
\eq\Delta\hat{\underline{v}}^q_l=-\mathrm{H}_{\hat{\underline{v}}_l^q}^{-1}\nabla
G_{\hat{\underline{v}}_l^q}\eeq where\eqn \nabla
G_{\hat{\underline{v}}_l^q}&=& qc-\left[%
\begin{array}{c}
  \tr(H(\hat{\underline{v}}_l^q)^{-1}Y_1) \\
  \vdots \\
  \tr(H(\hat{\underline{v}}_l^q)^{-1}Y_{n^2-m}) \\
  \tr(H(\hat{\underline{v}}_l^q)^{-1}X_1) \\
\end{array}%
\right]\nn\\
\mathrm{H}_{\hat{\underline{v}}_l^q}&=&\\
&&\hspace{-22mm}\left[%
\begin{array}{ccl}
 \tr(H (\hat{\underline{v}}_l^q)^{-1}Y_1 H(\hat{\underline{v}}_l^q)^{-1}Y_1)  & \hspace{-2mm}\tr(H (\hat{\underline{v}}_l^q)^{-1}Y_1 H(\hat{\underline{v}}_l^q)^{-1}Y_2) & \hspace{-2mm}\ldots   \\
\tr(H (\hat{\underline{v}}_l^q)^{-1}Y_2 H(\hat{\underline{v}}_l^q)^{-1}Y_1)        &  \hspace{-2mm}\tr(H (\hat{\underline{v}}_l^q)^{-1}Y_2 H(\hat{\underline{v}}_l^q)^{-1}Y_2) &    \\
 \vdots &  & \hspace{-2mm}\ddots    \\
\end{array}%
\hspace{0mm}\right]\nn \\
\eeqn
are the gradient and the Hessian computed at
$\hat{\underline{v}}_l^q$, respectively. Note that, it is not
difficult to prove that \begin{enumerate}
\item The sequence $\{\underline{v}^q_l\}_{l\geq 0}$ generated by
the algorithm is contained in the compact set
$\Nc=\Set{\underline{v}\in\mathrm{int}(\Ic)\;|\;
    -\frac{1}{\sqrt n}\leq c^T\underline{v}\leq c^T\underline{v}_0^q}$;
    \item $g_q$ is twice differentiable and strongly convex on $\Nc$;
    \item The Hessian of $g_q$ is {\em Lipschitz} continuous on
    $\Nc$.
\end{enumerate}
Accordingly, the Newton algorithm with backtracking globally
converges, \cite[9.5.3]{BOYD_CONVEX_OPTIMIZATION}. Moreover, the
rate of convergence is quadratic in the last stage. Finally, note
that the found solution $\hat{\underline{v}}^q$ satisfies the
following inequalities, \cite[p. 566]{BOYD_CONVEX_OPTIMIZATION},
\eq \label{convergenza_sol_barrier} c^T \underline{v}^\circ\leq
c^T \underline{v}^q\leq c^T \underline{v}^\circ+\frac{n}{q}\eeq
where $\underline{v}^\circ$ is a solution to Problem
\ref{min_eigen_pb} and $\frac{n}{q}$ the accuracy of
$c^T\underline v^q$ with respect to the optimal value
$c^T\underline{v}^\circ$.}

This method works well only setting a moderate accuracy. To
improve the accuracy, we can iterate the above Newton algorithm
and in each iteration we gradually increase q in order to find a
solution $\underline v^\xi$ with a specified accuracy $\xi>0$
\cite[p. 569]{BOYD_CONVEX_OPTIMIZATION}:
\begin{enumerate}
\item    Set the initial conditions $q_0>0$ and $\underline{v}^{q_0}=\left[%
\begin{array}{cccc}
   0 &\ldots & 0& \sqrt n(-\lambda_{\mathrm{min}}(\tilde\rho_0)+1)  \\
\end{array}%
\right]^T\in \mathrm{int}(\Ic)$.
\item At the $k$-th iteration compute $\underline v^{q_k}\in \mathrm{int}(\Ic)$ by minimizing $g_{q_k}$ with starting
point $ \underline v^{q_{k-1}}$ by using the Newton method
previously presented.
\item Set $q_{k+1}=\beta q_{k}$ with $\beta>1$ closer to one.
\item Repeat steps 2 and 3 until the   condition  $ \frac{n}{q_k}< \xi$  is satisfied.
 \end{enumerate}

Finally, we deal with the problem to compute a solution to Problem
\ref{prob_ricerca_rho} with kernel $\Kc$ when $\mu=0$. In this
situation, consider the non-empty convex set \eq
\Ic_*=\Set{\underline{w}\;|\; \tilde \rho_0+\sum_{i=1}^{n^2-m}w_i
Y_i\geq 0},\eeq where $\underline{w}=\left[\begin{array}{ccc}
                             w_1 & \ldots & w_{n^2-m} \\
                           \end{array}
                         \right]^T\in\Rs^{n^2-m}
$. Thus, the set of solutions to Problem \ref{prob_ricerca_rho} is
\eq \Sc=\Set{\tilde\rho_0+\sum_{i=1}^{n^2-m}w_i Y_i\;|\;
\underline{w}\in\Ic_*}.\eeq We wish to compute a matrix
$\rho^\circ\in\Sc$ having kernel corresponding to the minimum common kernel $\Kc$. To this aim consider the
following problem. \pb \label{problema_rho_perturbato}Pick
$\underline{u}\in \Rs^{n^2-m}$ at random and solve:\eq
\underline{w}^u=\arg\min_{\underline{w}\in\Ic_*}(\underline{w}-\underline{u})^T(\underline{w}-\underline{u})
.\eeq\epb Note that:
\begin{itemize}
  \item $\Ic_*$ is compact (i.e. closed and bounded)
  \item $(\underline{w}-\underline{u})^T (\underline{w}-\underline{u})$ is continuous and strictly
  convex on $\Ic_*$.
\end{itemize}
By Weiestrass' theorem, the above problem admits a unique solution
$\underline{w}^u$.

Let us define $\rho^{\circ,u}:=\tilde\rho_0+\sum_{k=1}^{n^2-m}w_i^u
Y_i\in\Sc$ and $\rho^u:=\tilde\rho_0+\sum_{k=1}^{n^2-m}u_i
Y_i\in\Sc$. It is easy to see that Problem
\ref{problema_rho_perturbato} can be rewritten in terms of the
matrix $\rho^{\circ,u}$ as follows:  \eq
\rho^{\circ,u}=\arg\min_{\rho\in\Sc}\|\rho^u-\rho\|^2_F,\eeq where
$\|\cdot\|_F$ denotes the {\em Frobenius} matrix norm. Thus,
$\rho^{\circ,u}$ is the closest matrix in $\Sc$ (with respect to the
{\em Frobenius} norm) to the matrix $\rho^{u}$ belonging to the
hyperplane characterized by the constraints
(\ref{constraint_per_rho}). Hence, randomly generating a finite
sequence $\Set{\underline{u}_1\ldots \underline{u}_l}$ of elements
in $\Rs^{n^2-m}$ we obtain a subset
$\Uc:=\Set{\rho^{\circ,u_1}\ldots \rho^{\circ,u_l}}$ contained in
$\Sc$.
Construct the convex combination $\bar\rho:=\frac{1}{l}\sum_{j=1}^l\rho^{\circ,u_j}.$ Then $\bar\rho$ has minimal kernel if either one of the $\rho^{\circ,u_j}$ belongs to the interior of ${\cal S}$, or $\rho^{\circ,u_j}$ belong to different faces of ${\cal S},$ which is a compact convex set. By the randomized construction, the probability of remaining on the boundary of ${\cal S}$  becomes small as $l$ grows.

Concerning the computation of $\underline{w}^u$, we can resort a
Newton-type algorithm  with logarithmic barrier named {\em
exterior-point method}, \cite[Chapter 4]{FIIACCO_MCCORMIK}.

\subsection{Continuity of $\lambda^\circ$ with respect to $\bar
f$}\label{continuity_lambda0} We show that the solution
$\lambda^\circ$ is continuous with respect to the data set
$\bar{f}$. To this aim we take into account the following result,
see \cite[Theorem 3.1]{RAMPONI_WELL_POSEDNESS_2010}. \teo Let $A$
be an open and convex subset of a finite-dimensional euclidean
space $V$. Let $h:A\rightarrow \Rs$ be a strictly convex function,
and suppose that a minimum point $\bar{x}$ of $h$ exists. Then,
for all $\varepsilon>0$, there exists $\delta>0$ such that, for
$p\in\Rs^n$, $\|p\|<\delta$, the function $h_p:A\rightarrow \Rs$
defined as \eq h_p(x):=h(x)-\Sp{p}{x}\eeq admits a unique minimum
point $\bar{x}_p$, and moreover \eq
\|\bar{x}_p-\bar{x}\|<\varepsilon.\eeq \eteo Consider \eq
J(\lambda,\bar{f})=\tr(\e{\log\tau-I -L^*(\lambda)})+\Sp{\lambda}
{\bar{f}}\eeq where we make the dependence of $J$ upon $\bar{f}$.
Then, the unique minimum point is
\eq\lambda(\bar{f})=\arg\min_{\lambda\in\Rs^m}
J(\lambda,\bar{f}).\eeq Let $\delta f\in\Rs^m$ be a perturbation
of $\bar{f}$. We have $J(\lambda,\bar{f}+\delta
f)=J(\lambda,\bar{f})+\lambda^T \delta f $. Applying the previous
theorem, where $\delta f $ is $-p$, we have: $\forall \;
\varepsilon>0$ $\exists \;\delta >0$ s.t. if $\|\delta f\|<\delta$
then $J(\lambda,\bar f+\delta f)$ admits a unique minimum point
\eq\lambda(\bar{f}+\delta f)=\arg\min_{\lambda\in\Rs^m}
J(\lambda,\bar{f}+\delta f)\eeq and \eq \|\lambda(\bar{f}+\delta
f)-\lambda(\bar{f})\|<\varepsilon.\eeq Accordingly, the map
$\bar{f}\mapsto \lambda(\bar{f})$ is continuous.

\bibliographystyle{IEEEtran}
\bibliography{biblio_spectral-1-1-2,bib-francesco-3-1}

\end{document}